# On the Separating Flow Behind a Cylinder: Insights from the Principle of Minimum Pressure Gradient

Mohamed Shorbagy & Haithem Taha

January 3, 2025

**Abstract**

In this paper, we study the free streamline theory for the separating flow over a circular cylinder. The objective of this paper is twofold: (i) to demonstrate the validity of the condition of matching curvature, and (ii) to obtain a reasonable estimate of the separation angle in the subcritical regime ($Re = 10^4 - 10^5$) without explicitly modeling the boundary layer. For the former goal, we study Roshko's free streamline model [Ros54]; it is an ideal flow model with sheets of discontinuities that represent the separating shear layers in the near wake region. It is known that the model fails to predict the correct separation angle over a curved surface. Roshko attributed this discrepancy to the condition of matching curvature (known as the condition of smooth detachment), which asserts that the curvature of the separating streamline at the separation point must match that of the cylinder. We show that the condition of matching curvature is legitimate and is not the real culprit for the failure of Roshko's model in predicting the correct separation angle. We also show that Roshko's model is non-unique. There are solutions in Roshko's family with matching curvature whose separation point matches experimental measurements. As for the second goal, we rely on the principle of minimum pressure gradient (PMPG), which asserts that, an incompressible flow evolves from one instant to another in order to minimize the total magnitude of the pressure gradient over the domain. Encouraged by the fact that the flow characteristics in the range $Re = 10^4 - 10^5$ are fairly independent of $Re$, we aim to predict the separation angle in this regime without modeling the boundary layer—a task that may seem impossible, though anticipated by Prandtl in his seminal paper [Pra04]. Over the family of kinematically-admissible, equilibrium flows, we utilize the PMPG to single out the separating flow with the minimum pressure gradient cost. Interestingly, the obtained separation angles match the experimental measurements over the regime $Re = 10^4 - 10^5$.

## 1 Introduction

In response to d'Alembert's paradox—which asserts that in an ideal potential flow, the pressure forces surrounding an object are perfectly balanced, resulting in zero drag [Rob65, Ste81]—Helmholtz and Kirchhoff introduced the free streamline theory as a resolution [Kir69]. The fundamental concept behind this theory is that the flow separates from the surface forming a wake behind it in the form of sheets of discontinuity. This discontinuity in the velocity field vitiates most of the conservation laws commonly known in ideal fluids (energy, vorticity, circulation, etc), as shown by Eyink [Eyi06, Eyi08, Eyi21, Eyi24]. In particular, it leads to pressure drag, even in an ideal fluid [HBL96], which is distinct from friction drag caused by shear forces acting on the surface. The principal objectives of the theory are: (i) to determine the "free streamlines" (i.e., sheets of discontinuity) that define the wake, beyond which the flow remains potential, and (ii) to compute the resulting pressure drag. This theory assumes that the pressure, and consequently the velocity, along these free streamlines remains constant—a notion that aligns with experimental observations [FJ27, Lam24, Ste81, Ros54, Ria21, Bro23]. The resulting velocity field is analytic everywhere except across the sheets of discontinuity. So, it is understood as a *weak* solution of Euler [Eyi08, Eyi21], which is expected to match the Navier-Stokes' limit as Reynolds number goes to infinity [Wu72], specifically in the near wake region. It is understood that, further downstream of the body, the sheets of discontinuity will become unstable, rolling up into a Von Karman street.

[LC07] and [Vil11] have made considerable advances to the mathematical foundations of the theory, followed by the efforts of [Wei23, Wei24a, Wei24b, Wei26] and [LW34, Ler35] proved the existence of solutions for specific shapes and demonstrated uniqueness for certain classes of geometries.

The development of this theory has generally been non-constructive [Zar52], and only a limited number of shapes have been addressed. For planar barriers, a solution can be directly obtained using the Schwarz-Christoffel transformation with the aid of the powerful hodograph method [MT69]. In contrast, for curved barriers, [Gre16] suggested initiating the solution by *"writing down a likely expression ... and then investigating the streaming* *motion implied and the shape of the boundary"*. Following Greenhill's suggestion and using Levi-Civita's transformation [LC07], [Bro23] proposed an approximate solution for circular and elliptical cylinders, which can achieve any desired level of accuracy by including additional terms in a series expansion. For a circular cylinder, the separation angle was predicted to be 55°, based on the assumption that the curvature of the free streamline at the separation point matches the curvature of the cylinder, leading to smooth separation. This condition is sometimes referred to as the condition of "smooth detachment" [Wu72]. This assumption also known as the Villat-Brillouin condition, traces back to [Bri11] and [Vil14].

It should be noted that the Helmholtz-Kirchhoff free streamline theory has predicted drag coefficients significantly lower than experimental values. This discrepancy arises because the theory assumes that the velocity along the free streamline is not only constant but fixed at the free stream value, contrary to experimental evidence, where the velocity is found to be significantly higher than the free stream. Several models have been proposed to extend the free streamline theory to allow for an arbitrary value of the constant speed along the free streamline. These extensions include Riabouchinsky's image model for the separating flow behind a flat plate [Ria21], where an image flat plate is introduced to close the wake. An alternative approach (the open wake model) was first introduced by [Jou90] and independently rediscovered by [MY53], [Ros54], [Epp54], and later extended by [Wu62] to address lifting flows past arbitrary bodies. There is also the Re-entrant Jet Model [Gil60a] and the Cusped Wake Model [Bri11] among other approaches. These models are summarized in a review article by [Wu72]. Since then, the theory has been continuously (though very slowly) developing, see the recent articles by [MM24] and [MDM24].

In this work, we will give focus to one of these classical extensions, specifically Roshko's model [Ros54], where the velocity along the separating streamline (i.e., the free streamline) is adjusted to $kU_\infty$, where $k >= 1$ is determined from experimental measurements. The global picture of the flow field resulting from Roshko's model in the case of a flat plate was in excellent agreement with experimental measurements. However, for circular cylinders, there remained a significant discrepancy in the location of the separation point. Roshko attributed this discrepancy to the use of the matching curvature condition, i.e., the Villat-Brillouin condition.

In this work, we investigate the validity of the matching curvature condition in Roshko's study [Ros54], examining whether it is indeed the cause of the discrepancy observed in the location of the separation point. If not, we propose an alternative explanation for this discrepancy.

Furthermore, the fact that the characteristics of the separating flow are fairly independent of Reynolds number over the subcritical regime $Re = 10^4 - 10^5$ [Zdr97] may suggest developing a low-fidelity approach that can approximately predict the separation point in the subcritical regime without explicitly accounting for the flow details inside the boundary layer. This endeavor matches Prandtl's visionary statement in his seminal paper on the concept of boundary layer [Pra04]:

*"The most important practical result of these investigations is that, in certain cases, the flow separates from the surface at a point entirely determined by external conditions"*.

To achieve this goal, we rely on the Principle of Minimum Pressure Gradient (PMPG) developed by the authors [TGS23]. That is, among all kinematically-admissible equilibrium solutions of Euler, the actual separating flow is the one that minimizes the total magnitude of the pressure gradient force in the domain.

During the course, we will show that Roshko's model is not robust enough to perform minimization of the pressure gradient. Instead, we found the recent model of [MM24] more suitable for integration with the PMPG formulation due to its simplicity and robustness, despite its relatively lower accuracy in comparison to Roshko's.

## 2  Roshko's Model

The predicted drag by Kirchhoff's free streamline theory is significantly less than reality due its basic assumption: The velocity on the free streamline is assumed to be equal to the free stream velocity, which contradicts experimental observations. [Ros54] extended Kirchhoff's theory by constructing an ideal-fluid model for the separating flow behind an obstacle such that the free streamline velocity is $kU_\infty$, for an arbitrarily given value of $k \geq 1$. This extension leads to a base pressure coefficient that is less than zero is essential to agree with the experimental observations.

[Ros54] applied his extended theory to different geometries, including a flat plate, a wedge, and a circular cylinder. As pointed out by [Wu72], there is a fundamental difference between the first two cases (flat plate and wedge) and the last case (circle); in the former cases, the separation point is known a priori (at the sharp edges). In contrast, the location of the separation point is not trivially determined in the case of a circular cylinder, or any curved surface. Wu noted that separating flows with fixed separation tend to be steadier with a more uniform back pressure, in contrast to flows past curved bodies, which typically exhibit stronger oscillations. The reason

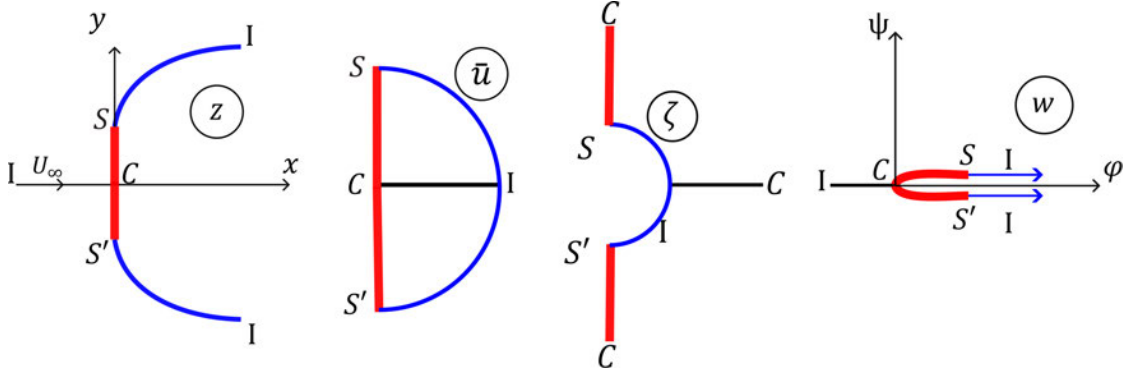

Figure 1: Conformal mapping of a flat plate normal to the free stream (the Kirchhoff problem), where $z$ is the actual flow domain, $\bar{u}$ is the hodograph plane (plane of complex velocity), $W$ is the complex potential plane, and $\zeta$ is the reciprocal of $\bar{u}$.

for this difference in behavior is attributed to the fact that a flow towards a sharp edge is always associated with a favorable pressure gradient (hence accelerating); and changes in Reynolds number cause little effects to this behavior. We focus in this paper on the latter case where significant discrepancies in the location of the separation point were found between experimental measurements and the extended theory of Roshko.

The typical approach in the free streamline theory is to utilize the knowledge of the general features of the flow in the plane of complex potential $W(z) = \phi + i\psi$ and in the hodograph plane—the plane of complex velocity $\bar{u} = \frac{dW}{dz} = qe^{-i\theta}$—to construct a conformal map between the two planes using the Schwarz-Christoffel transformation or the Levi-Civita method. For example, in the case of a flat plate normal to the free stream (the Kirchhoff problem), the boundary $I - S' - C - S - I$ of the actual flow domain (i.e., the $z$-plane) corresponds to a slit on the real axis (the $\phi$-axis or $\psi = 0$) in the $W$-plane because it is the same streamline, as shown in figure 1. On the other hand, since the direction of the flow (defined by the angle $\theta$) is known on $I - C$ and $C - S$, these boundaries would represent horizontal and vertical lines in the hodograph plane ($\bar{u}$-plane), as shown in the same figure. It is usually easier to consider the plane of $\zeta = \frac{1}{\bar{u}} = \frac{1}{q}e^{i\theta}$ instead of $\bar{u}$ since the argument of $\zeta$ directly represents the actual angle $\theta$ of the velocity vector. Moreover, since (according to the free streamline theory) the flow speed along the separating streamline $S - I$ is constant, then the boundary $S - I$ should represent a circular arc (i.e., a curve with constant radius) in the $\bar{u}$ or $\zeta$ planes. This formulation allowed construction of the boundaries in the $W$- and $\zeta$-planes. The goal is then to construct a conformal map between the corresponding boundaries in the two planes. Once the mapping $W(\zeta)$ is obtained, a differential equation for the flow velocity in the actual domain can be written as

$$\frac{d\zeta}{dz} = \frac{1}{\zeta \, dW/d\zeta},$$

where $z = x + iy$ is the complex coordinate in the actual plane. The solution of the above equation yields the reciprocal of the complex velocity $\bar{u}$ in the actual domain: $\zeta(z)$.

To construct the conformal map $W(\zeta)$, one may use the Schwarz-Christoffel transformation, which maps the boundary of any *simple* polygon to the real axis. Hence, an intermediate plane (say the $t$-plane), which has the boundary $I - S' - C - S - I$ on its real axis (from $-\infty$ to $\infty$), is needed. The Schwarz-Christoffel transformation can then be used to construct two maps $W(t)$ and $\zeta(t)$. Moreover, since the boundaries of the $\zeta$-plane do not constitute a polygon, it is more convenient to consider the logarithmic hodograph plane instead: $\Omega = \log \zeta = \log \frac{1}{q} + i\theta$. In the $\Omega$-plane, the real part corresponds to the velocity magnitude and the imaginary part exactly represents the angle $\theta$ of the velocity vector. As such, the domain becomes a polygon in the $\Omega$-plane, allowing the use of the Schwarz-Christoffel transformation to construct the maps $\Omega(t)$ and $W(t)$, from which $W(\zeta)$ can be inferred. This approach is detailed in several references [e.g., Lam24, MT69, Ros54, Gil60b, Fel19]. Another approach for constructing the map $W(\zeta)$ is the Levi-Civita method [Gol38], which transforms semi-circles into half planes. In this case, the boundary $I - S' - C - S - I$ becomes a semi-circle in the intermediate plane (say the $\tau$-plane), resulting in the maps

$$\sqrt{W} = \frac{\tau^2 - 1}{2i\tau} \quad \text{and} \quad \Omega = \log \frac{1 + \tau}{1 - \tau}. \tag{1}$$

This is the approach adopted by [Bro23] and [Ros54] in their models for the separating flow over a cylinder problem, shown in figure 2. However, the challenge with the circular cylinder, in contrast to the flat plate case, lies in the fact that the cylinder boundary $CS$ is not known in the $\zeta$ or $\Omega$ planes. To resolve this issue, Brodetsky added a correction $\Omega'(\tau)$ to the transformation of the flat plate case $\Omega(\tau)$, given in equation (1):

$$\Omega(\tau) = \log\frac{1+\tau}{1-\tau} + \Omega'(\tau), \tag{2}$$

where $\Omega'(\tau)$ is expressed in the series form:

$$\Omega' = A_1\tau + \frac{1}{3}A_3\tau^3 + \frac{1}{5}A_5\tau^5 + \ldots, \tag{3}$$

where the coefficients $A_1, A_3, \ldots$ are to be determined. Since $|\tau| \leq 1$, the effect of higher-order ones in $\tau$ is smaller than lower-order terms. In their solutions, Brodetsky and Roshko truncated the series after two terms.

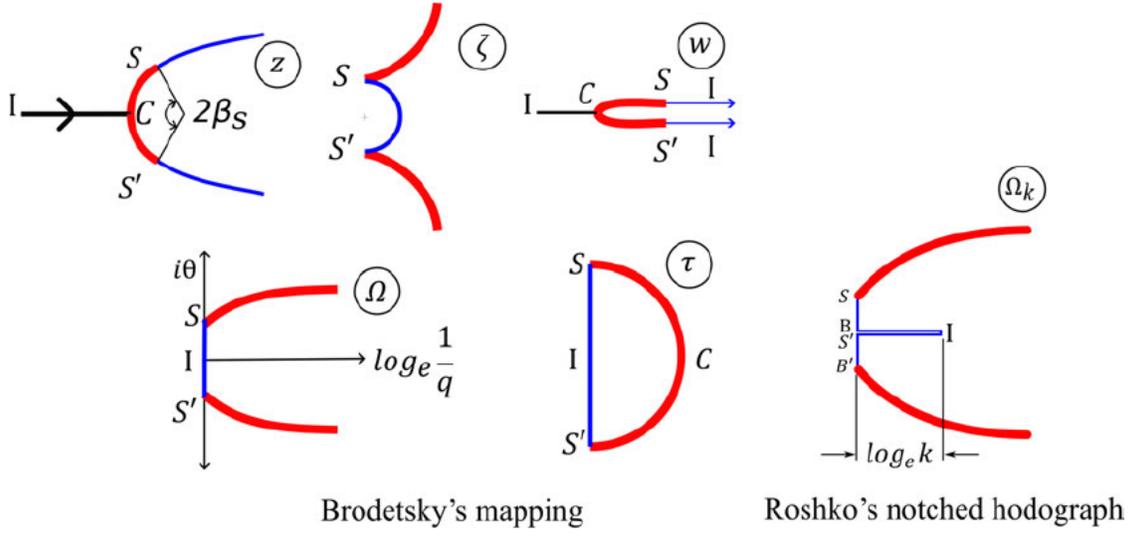

Figure 2: (a) Brodetsky's mapping of the separating flow over a cylinder using Levi-Civita method where $\tau$ is an intermediate plane in which the boundary $I - S' - C - S - I$ becomes a semi-circle. (b) Roshko's notched Hodograph allowing the free streamline to curve at a constant speed $kU_\infty$ until it levels out at point $B$, beyond which the flow decelerates horizontally until it reaches $U_\infty$ to respect the far field boundary condition.

Brodetsky adopted the Villat-Brillouin condition of smooth detachment, which asserts that the radius of curvature of the free streamline at the separation point must be equal to the radius of curvature of the cylinder. He derived the following expressions for the radius of curvature along the free streamline and along the cylinder:

$$-\frac{ds}{d\theta}\Big|_{FSL} = \frac{(1-\rho^4)/2\rho^3}{2/(1+\rho^2) + A_1 - A_3\rho^2 + A_5\rho^4 - \ldots}, \tag{4}$$

$$\frac{ds}{d\theta}\Big|_{Cyl} = \frac{\frac{2}{q}\sin\sigma}{A_1 + A_3\frac{\cos 3\sigma}{\cos\sigma} + A_5\frac{\cos 5\sigma}{\cos\sigma} + \ldots}, \tag{5}$$

where $\rho$ and $\sigma$ come from the definition $\tau \equiv \rho e^{i\sigma}$ with the separation point corresponding to $(\rho, \sigma) = (1, \pi/2)$, and

$$\frac{1}{q} = \sqrt{\frac{1+\cos\sigma}{1-\cos\sigma}} \cdot e^{A_1\cos\sigma + \frac{1}{3}A_3\cos 3\sigma + \ldots}$$

Equation (4) implies that the radius of curvature of the free streamline at the separation point ($\rho = 1$) is zero, resulting in an infinite curvature, which violates the Villat-Brillouin condition of smooth detachment. To enforce a finite curvature (i.e., non-zero radius of curvature), the denominator of equation (4) must be zero at $\rho = 1$:

$$1 + A_1 - A_3 + A_5 = 0. \tag{6}$$

Brodetsky showed that, once the condition (6) of finite radius of curvature is satisfied, the Villat-Brillouin condition of smooth detachment is naturally achieved; i.e., equation (6) does not only imply that the radius of curvature of the free streamline is finite at the separation point, but also equal to the radius of curvature of the cylinder at the same point: the limit of equation (5) as $\sigma \to \pi/2$ and $q \to 1$.

The coefficients $A$'s are determined such that $\frac{ds}{d\theta}|_{Cyl}$ matches the given curvature of the body at all points (i.e., for all $\sigma \in [0, \pi/2]$), while satisfying equation (6). That is, for a circular arc, the $A$'s are determined to make the radius of curvature in equation (5) approximately constant for all $\sigma \in [0, \pi/2]$ to any arbitrarily required accuracy. Brodetsky showed that it is sufficient to construct an almost circular arc using only two coefficients $A_1, A_3$ that satisfy equation (6). In this case, the variation of the radius of curvature was found to be below 3.5%.

[Ros54] extended Brodetsky's model to the general case of an arbitrarily given $k \geq 1$ using the notched hodograph, shown in figure 2 (b). The idea is that the free streamline curves at a constant speed $kU_\infty$ until it levels out at some point $B$, beyond which the flow decelerates horizontally until it reaches $U_\infty$ to respect the far field boundary condition. His notched logarithmic hodograph $\Omega_k$ is related to Brodetsky's hodograph $\Omega$ as:

$$\sinh^2 \Omega_k = \left(\frac{k^2+1}{2k}\right)^2 \left[\sinh^2 \Omega + \left(\frac{k^2-1}{k^2+1}\right)^2\right]. \tag{7}$$

Clearly, as $k \to 1$, we recover Brodetsky's hodograph: $\Omega_k \to \Omega$.

Following Brodetsky, Roshko considered two coefficients $A_1, A_3$, which are determined, for a given value of $k$, using the following iterative scheme:

1. Choose a value for $A_1$ and determine $A_3$ from equation (6) assuming the remaining $A$'s are zeros.

2. Along the cylinder's arc $CS$, $\tau = e^{i\sigma}$, where $\sigma$ goes from 0 to $\pi/2$. Using $A_1$ and $A_3$ from the previous step, $\Omega'$ and $\Omega$ can be obtained from equations (3) and (2), respectively.

3. $\Omega_k$ can then be computed from equation (7), resulting in $\zeta = e^{\Omega_k}$.

4. The complex potential $W$ can be computed from equation (1). In particular, its derivative, $\frac{dW}{d\tau} = \frac{1-\tau^4}{2\tau^3}$, can be computed.

5. The coordinate $z$ of the constructed shape corresponding to the assumed value of $A_1$ is computed as

$$z|_{Cyl} = \int \zeta(\tau) \frac{dW}{d\tau}(\tau) \, d\tau.$$

6. The radius of curvature is computed as

$$\frac{ds}{d\theta} = \frac{d|z_{Cyl}|}{d \arg(z_{Cyl})}.$$

7. If the maximum deviation of $\frac{ds}{d\theta}$ along the arc from its mean value is below a certain tolerance, then convergence is achieved. If not, $A_1$ needs to be updated, and the scheme is repeated from step (1).

Step (3) requires special attention, though not emphasized by Roshko. Equation (7) provides two solutions for

$$\Omega_k = \sinh^{-1}\left(\pm\sqrt{\text{RHS}}\right),$$

where RHS is the right-hand side of equation (7). Note that

$$\sinh \Omega_k = \sinh(R + i\theta) = \sinh R \cos \theta + i \cosh R \sin \theta,$$

where $R = \log\left(\frac{1}{q}\right)$. Since $\theta > 0$ in the whole domain, the imaginary part of $\sinh \Omega_k$ must be positive. Therefore, we pick the root which results in $\text{Im}(\sinh \Omega_k) > 0$. If the imaginary part is close to zero, which occurs on the $x$-axis and near infinity, it is expected that $\theta < \pi/2$ and $q < 1$ (i.e., $R > 0$) in this region. Hence, we pick the root that results in a positive real part of $\sinh \Omega_k$.

Once $A_1, A_3$ are determined, the whole flow field is obtained. In particular, the velocity field is determined at every point $\bar{u} = e^{-\Omega_k}$, from which the pressure distribution over the cylinder as well as in the domain is determined. Also, the separation angle can be determined: it is the argument of $z|_{Cyl}$ corresponding to $\tau = i$. So, according to Roshko, his theory has one free parameter ($k$), which must be determined from external considerations (experimental measurements or high-fidelity simulations).

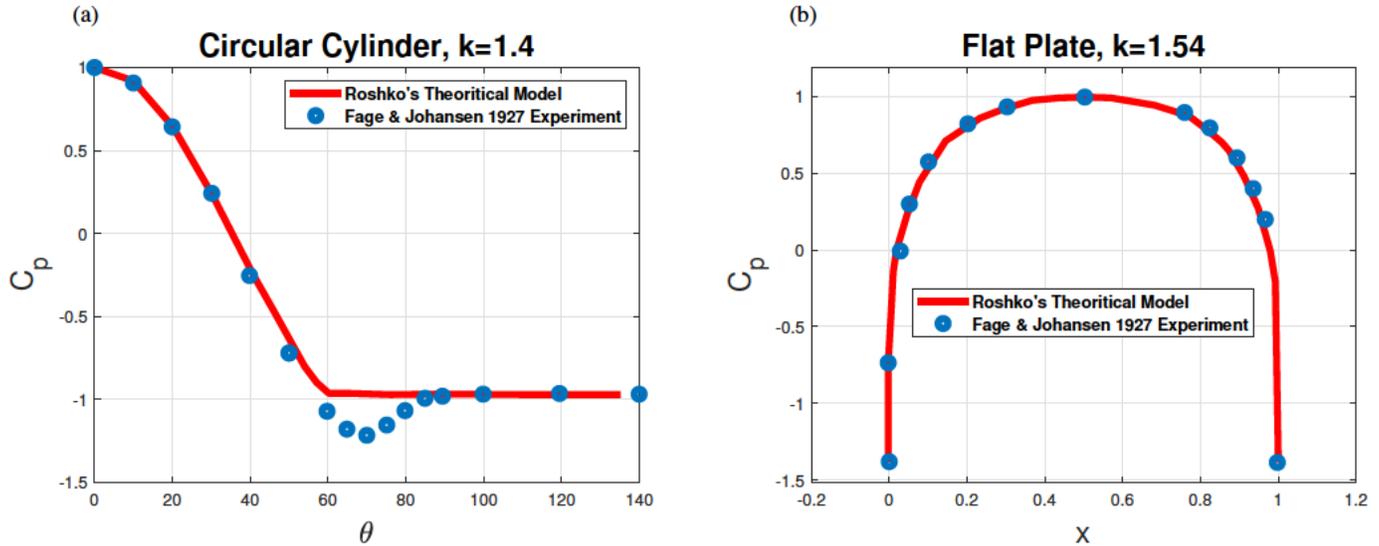

Figure 3: Comparison between Roshko's theoretical results of the pressure distribution on (a) circular cylinder, and (b) a flat plate, against the experimental measurements of [FJ27]. Perfect agreement is found for the flat plate case all over the plate's surface. In contrast, a considerable discrepancy is found for the cylinder case near the separation point.

Roshko considered the experimental data of [FJ27] who measured the pressure distribution over the cylinder at $Re = 14,500$. Their measurement of the wake pressure coefficient results in $k = 1.4$. Plugging this value of $k$ in Roshko's model, he obtained a full picture of the flow field. He predicted a separation angle of $62°$ in contrast to Brodetsky's result of $55°$ at $k = 1$. Roshko compared the pressure distribution over the cylinder resulting form his model to the measurements of [FJ27], see figure 3 (a).

The two distributions compare very well over the majority of the cylinder except near the separation point where Roshko's model predicted a favorable pressure gradient in contrast to the expected adverse pressure gradient commonly encountered near separation. In fact, Roshko's prediction ($62°$) of the separation angle is considerably different from the value ($82°$) found by [FJ27]. Since the theory returned accurate results in the flat plate case, as shown in figure 3 (b), Roshko attributed the discrepancy in the value of the separation angle and the pressure distribution near separation to the condition of matching curvature , presented in equation (6); he wrote "*Clearly the assumption that the streamline has the same curvature of the cylinder is not satisfactory. With this assumption, there is no adverse pressure gradient, whereas it is well-known that to separate boundary layer on a continuous surface an adverse gradient is necessary.*" [Ros54]

In this paper, we investigate Roshko's hypothesis, whether the condition of matching curvature (smooth detachment) is the real culprit behind the discrepancy found in the predicted separation angle.

## 3  Inspection of the Condition of Matching Curvature

To investigate Roshko's hypothesis—that the condition of matching curvature is the real culprit for the discrepancy in the separation angle—we begin our analysis by relaxing this condition expressed in equation (6). As such, if we truncate the series (3) after two terms similar to Brodetsky and Roshko, we will have two free parameters: $A_1, A_3$; no constraint is imposed to determine one of them from the other. This may imply non-uniqueness. We can have infinitely many combinations of $(A_1, A_3)$ that construct a circular shape with an acceptable accuracy; the matching condition picks only one of them.

The non-uniqueness issue is accentuated further with increasing the number of coefficients in Brodetsky's series. Adding more coefficients is known to enhance accuracy and reduce variation in the shape's radius, thereby providing a better approximation of a circular cylinder. However, it also introduces more free parameters, thereby expanding the set of candidate separating flows. To demonstrate this behavior, we consider the simple case of $k = 1$ (i.e., Brodetsky's), but with five coefficients: $A_1, A_3, A_5, A_7$, and $A_9$. We iteratively find a combination of these five coefficients such that the maximum deviation of the radius from the mean value is less than $1\%$.

Table 1: Multiple solutions from Roshko's model at $k = 1$ using five coefficients in the series (3) while relaxing the condition (6) for matching curvature. Multiple solutions can be constructed that match the cylinder surface very well, but with different flow configurations and separation angles, implying non-uniqueness of Roshko's model.

|  | A1 | A3 | A5 | A7 | A9 | $\beta_s^\circ$ |
|---|---|---|---|---|---|---|
| **Solution 1** | -1.0477 | 0.0716 | -0.0109 | -0.0044 | 0.0023 | 61.7 |
| **Solution 2** | -1.1372 | 0.0913 | -0.0223 | 0.0072 | 0.0044 | 67.0 |
| **Solution 3** | -1.2186 | 0.1108 | -0.0191 | -0.0040 | 5.47e-4 | 72.5 |
| **Solution 4** | -1.2547 | 0.1194 | -0.0259 | -0.0039 | 0.0053 | 74.8 |
| **Solution 5** | -1.3418 | 0.1412 | -0.0366 | 0.0073 | 0.0053 | 80.5 |

Table 2: Multiple solutions from Roshko's model at $k = 1.4$ using 12 coefficients in the series (3) while enforcing the condition (6) for matching curvature. Multiple solutions can be constructed that match the cylinder surface to arbitrary accuracy, but with different flow configurations and separation angles, implying non-uniqueness of Roshko's model even when enforcing matching curvature.

|  | A1 | A3 | A5 | A7 | A9 | A11 | A21 | $\beta_s^\circ$ |
|---|---|---|---|---|---|---|---|---|
| **Solution 1** | -0.9827 | 0.0062 | -0.0147 | -0.0118 | -0.0084 | 0 | -0.0002 | 62.1545 |
| **Solution 2** | -1.0902 | 0.0103 | -0.0087 | -0.0175 | -0.0092 | 0 | -0.1009 | 70.1453 |
| **Solution 3** | -1.1531 | 0.0062 | 0.0010 | -0.0265 | -0.0101 | 0 | -0.1419 | 75.2848 |
| **Solution 4** | -1.2088 | -0.0045 | 0.0139 | -0.0339 | 0.0012 | -0.0123 | -0.1430 | 80.0000 |
| **Solution 5** | -1.2448 | -0.0195 | 0.0357 | -0.0511 | 0.0076 | -0.0160 | -0.1149 | 83.4636 |

The results are presented in figure 4, where the corresponding values of the coefficients are:

$$A_1 = -1.2993, A_3 = 0.1294, A_5 = -0.0332, A_7 = 0.0099, A_9 = 0.0046$$

These values do not satisfy equation (6). Consequently, the corresponding flow does not satisfy the condition of matching curvature, as can be seen in figure 4 (a): the separating streamline (blue curve) has the same slope as the cylinder (red curve), but not the same curvature. In fact, mathematically speaking, the free streamline has an infinite curvature at the separating point in this case. This significant jump in curvature can be clearly seen in the peak of acceleration magnitude at the separation point, shown in figure 4(b).

Figure 4 (c) presents the resulting flow field where the separation angle is found to be $\beta_s = 77°$, which is quite different from Brodetsky-Roshko's value of 55° at $k = 1$. An interesting aspect of this result is that it represents only one of several possible solutions, all corresponding to the same value of $k$, but with different separation angles $\beta_s$. This non-uniqueness is demonstrated in Table 1, which presents a subset of these possible solutions for $k = 1$.

Clearly, this non-uniqueness is not confined to the case of $k = 1$ nor is it related to the relaxation of the matching curvature condition. For example, consider Roshko's case of $k = 1.4$ enforcing the condition (6) for matching curvature, but using 12 coefficients in Brodetsky's series for better accuracy. Note that as $k$ increases, more coefficients are required to maintain the same level of accuracy, as observed by Roshko; the largest value of $k$ he could analyze using two terms was 1.6. In this case, there are 11 free parameters, with one of the 12 coefficients, say $A_{23}$, determined in terms of the other 11 using equation (6).

Multiple solutions are obtained, each resulting in a different value of $\beta_s$, as shown in Table 2, which presents some of these solutions. Indeed, Roshko's model is non-unique even with enforcing the condition of matching curvature, which refutes Roshko's claim: *" if it be assumed that the streamline at separation has the same curvature as the cylinder, then there will be a unique value of $\beta_s$ for every value of k."* [Ros54]. The above results imply that Roshko's model has at least two free parameters ($k$ and $\beta$), and not only $k$ as posited by Roshko. In addition, the reported solution of Roshko at $k = 1.4$ that resulted in a separation angle of $\beta_s = 62^o$ (which did not match experimental measurements) is only one solution among infinitely many possible candidates. Moreover, some of these candidates better match the experimental measurements of [FJ27] (in terms of the separation angle and the pressure distribution near separation) than Roshko's, as discussed below.

The pressure distributions corresponding to the candidate solutions in Table 2 are shown in figure 5 (a), in comparison to the experimental measurements of Fage and Johansen. In particular, one candidate (Solution

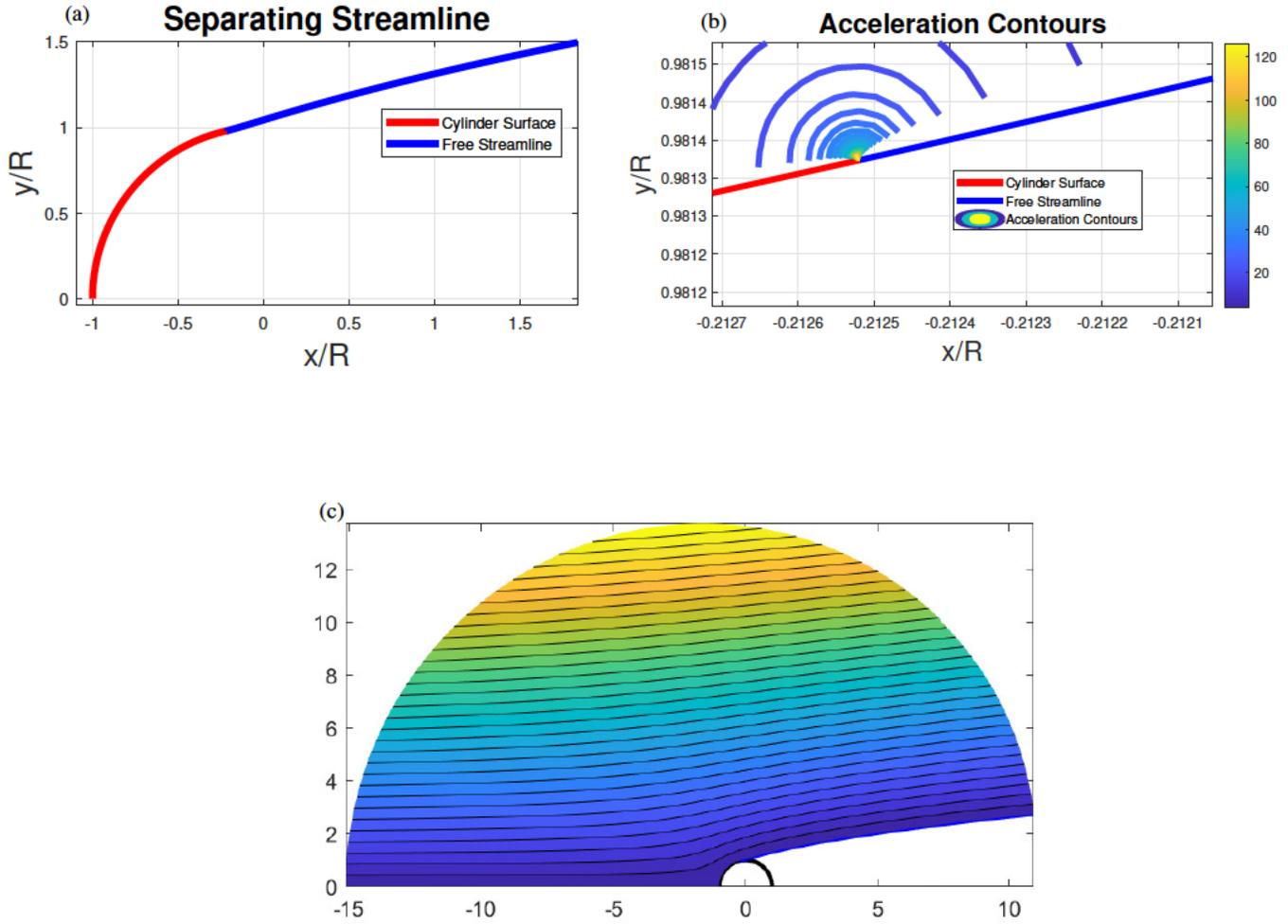

Figure 4: Results of Roshko's model at $k = 1$ using five coefficients of the series (3) without enforcing the matching curvature condition, showing non-uniqueness of Roshko's model. (a) Free streamline and cylinder's surface. The free streamline (blue curve) has the same slope as the cylinder (red curve), but not the same curvature. (b) Contours of acceleration magnitude, showing a significant jump at the separation point due to the difference in curvature of the free streamline and the cylinder's surface. (c) Streamline contours of the resulting flow field whose separation angle is $\beta_s = 77°$, which is significantly different from Roshko-Brodetsky's value of $55°$ at $k = 1$.
8

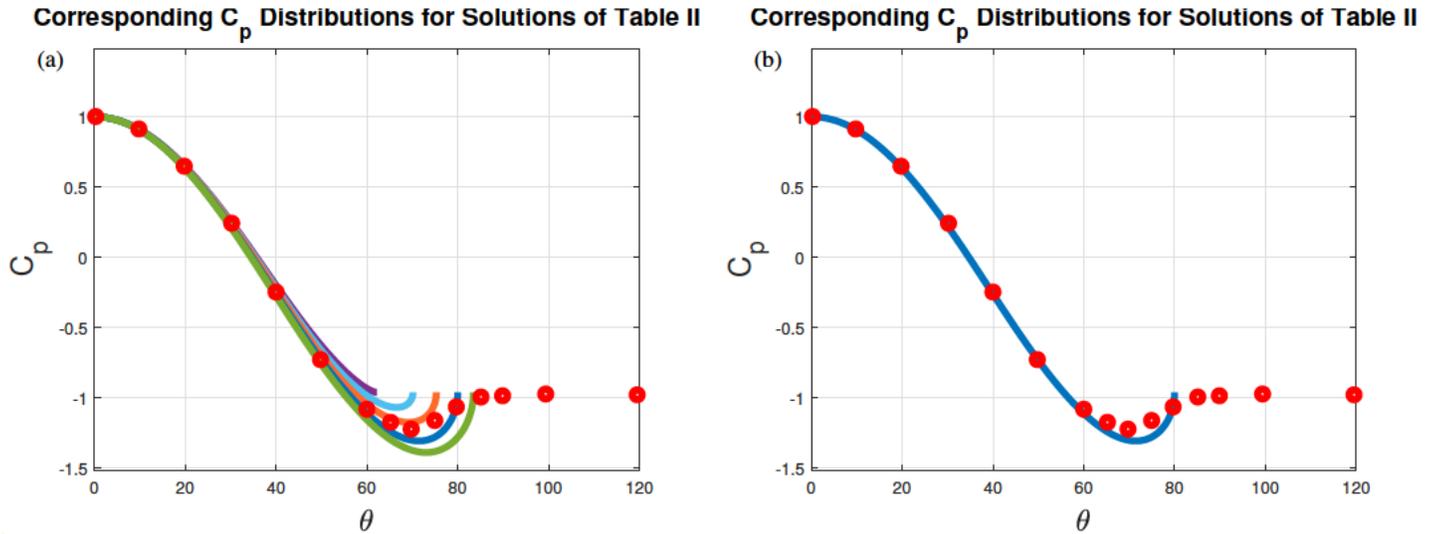

Figure 5: The pressure distributions of the candidate solutions from Roshko's model at $k = 1.4$, presented in Table 2, illustrating the non-uniqueness of Roshko's model, even with enforcing the condition of matching curvature. (a) The pressure distributions corresponding to the candidate solutions compared to the experimental measurements of [FJ27]. (b) The pressure distribution of one candidate (Solution 4), resulting in a separation angle $\beta_s = 80°$, matching the experimental value and almost perfectly aligning with the experimental measurements of [FJ27], demonstrating an adverse pressure gradient near separation.

4) results in a separation angle of $\beta_s = 80^o$, matching the experimental value. Moreover, the corresponding pressure distribution almost perfectly aligns with the experimental measurements, as shown in figure 5 (b). The candidate solution shows an adverse pressure gradient near separation, demonstrating excellent agreement with the experiment. This finding is related to the open problem in the mathematical theory of fluid flows about the convergence of Navier-Stokes' solutions to Euler's as the Reynolds number goes to infinity [CKV15, Mas07, Eyi08, Eyi21, Eyi24]. This family of solutions (e.g., Solution 4) consists of purely inviscid discontinuous (*weak*) solutions of Euler's equation whose members are capable of matching the zero-viscosity limit of Navier-Stokes. However, it remains to find a selection criterion that picks the right candidate from the family that matches the infinite-Reynolds number limit [Eyi24]. It is important to emphasize that the solutions presented in Table 2 excluding Roshko's candidate (Solution 1), exhibit an adverse pressure gradient near separation, as shown in figure 5. Given that these solutions satisfy the matching curvature condition, Roshko's assertion "*Clearly the assumption that the streamline has the same curvature as the cylinder is not satisfactory. With this assumption, there is no adverse pressure gradient*" [Ros54]—can therefore be refuted.

We have demonstrated that the discrepancy in the separation point between Roshko's model and experimental measurements cannot be attributed to the condition of matching curvature. Rather, it is in fact due to Roshko's choice of a specific solution from the family (by truncating the series after two terms). However, the validity of the matching curvature condition remains undecided at this stage. Interestingly this point could be studied using *The Principle of Minimum Pressure Gradient* (PMPG), as shown below.

## 4  Insights From the Principle of Minimum Pressure Gradient

[TGS23] extended Gauss' principle of least constraint [Pap14, KU01] from particle mechanics to incompressible flows, relying on the fact that for such flows, pressure is the Lagrange multiplier that enforces continuity [GS87, MK80, MAP20]. That is, in the language of analytical mechanics [Pap14, KU01, Lan12], the pressure is a *constraint force*. Hence, according to Gauss' principle, the total magnitude of the pressure force over the domain must be minimum at every instant, which is referred to as *The Principle of Minimum Pressure Gradient* [TGS23]. We proved that if the flow acceleration $u_t$ minimizes the cost functional

$$\mathcal{A} = \frac{1}{2} \int_{\mathcal{D}} \rho \left( u_t + u \cdot \nabla u - \frac{1}{\rho} \nabla \cdot \tau \right)^2 dx, \qquad (8)$$

at each time instant subject to the continuity constraint and no-penetration boundary condition, then it must satisfy the Navier-Stokes equation (see Theorem 1 in [TGS23]). That is, Navier-Stokes' equation is the first-order necessary condition for minimizing the pressure gradient cost (8). As such, the PMPG turns a fluid mechanics problem into a minimization one since we can focus only on minimizing the cost; and the resulting solution is guaranteed to satisfy Navier-Stokes'. For example, [AET24] utilized the PMPG formulation to solve the flow over a cylinder by performing pure minimization using physics-informed neural networks [DPT94, RPK19, LMMK21, DZ21]. [AD24] independently proposed a similar formulation to solve the lid-driven cavity problem. The reader is cautioned that Gauss' principle and the PMPG differ fundamentally from most variational principles in classical physics, which are typically derived from the principle of least action. In the PMPG, variations are taken with respect to $u_t$ as a function of space at a fixed time instant, see [TGS23] for more details on this point.

In the special case where no impressed forces (e.g., viscous forces) are applied, Gauss's principle—and its fluid mechanics analogue the PMPG—reduce to Hertz's principle of least curvature [Pap14]. The cost $\mathcal{A}$ then reduces to the *Appellian*

$$S = \frac{1}{2} \int_{\mathcal{D}} \rho \left( u_t + u \cdot \nabla u \right)^2 dx. \tag{9}$$

Minimizing the Appellian $S$ (in the absence of applied forces) is equivalent to minimizing the total curvature in the system [Pap14]—hence the name least curvature.

The PMPG and its special case of least curvature were successfully applied to various problems in incompressible fluid mechanics, such as the Stokes' second problem [TGS23], the lid-driven cavity [AD24], the flow over a cylinder [AET24]. Of particular interest is the application of the PMPG to the problems of lifting airfoils [GT22, TG23] and rotating cylinders [ST24]. In these two problems, it was deemed impossible to obtain a satisfactory solution without solving Navier-Stokes' equation or modeling the boundary layer dynamics. However, the PMPG allowed selecting the unique solution from Euler's family that matches the expected Navier-Stokes' limit. For example, the least-curvature solution of the flow over a sharp-edged airfoil [GT22, TG23] recovered the well-known limiting solution of Kutta; and the same principle/criterion extended Kutta's theory to smooth shapes that do not have sharp edges. Also, in the case of a rotating cylinder, the PMPG solution [ST24] perfectly matched the zero-viscosity limit of Glauert's solution of Prandtl's boundary layer equations [Gla57].

Having provided some background on the PMPG and least-curvature, we may now proceed to justify the condition of matching curvature (6) on physical grounds. Recall that if the condition (6) is not satisfied, the radius of curvature of the free streamline at the separation point will be zero, resulting in infinite curvature at that point—a behavior demonstrated by a large peak in acceleration near separation, as shown in figure 4 (b). This very large peak in the local acceleration is expected to result in a very large total cost (9). Hence, the optimizer is expected to naturally converge to a solution that satisfies the condition (6). Therefore, the matching curvature condition, *i.e.*, ensuring that the curvature of the free streamline matches that of the cylinder at the separation point, is then fundamentally driven by a physical principle: the principle of least curvature.

The above argument provides a qualitative justification of the condition of matching curvature. However, one may ask the following legitimate question: What does the least-curvature solution from Roshko's family look like? Can we find the right solution from Roshko's family (which matches experimental results) by minimizing the pressure gradient—equivalently the curvature cost (9)? Unfortunately, when considering multiple terms in the series, Roshko's model becomes inappropriate for numerical optimization; it is highly sensitive to changes in the free parameters (coefficients of the series). For example, changing the coefficients of Solution 4 in Table 2 by 5%, leads to a substantial impact: the cylinder radius variation jumps from 1% to 12% and the separation angle $\beta_s$ shifts from 80° to 90°.

Another important issue in Roshko's model, when it comes to minimizing acceleration or curvature, is the notch he introduced in the hodograph plane to allow a free streamline speed larger than $U_\infty$. In his work, [Ros54] showed a smoother way (*"elliptic hodograph"*) to achieve the same goal, but he opted for the notched hodograph. The notch induces a fictitious jump in the acceleration at the point where the free streamline becomes horizontal (i.e., parallel to the free stream); i.e., point "B" in figure 2 (b). An example of this jump is clearly seen in the contours of acceleration magnitude presented in figure 6 (a) for Roshko's flow (corresponding to $\beta_s = 62°$). The region around the point $B$ experiences the largest acceleration in the whole field. While it is true that the flow starts to decelerate at this point, it is however nonphysical to do it in such an abrupt manner that imposes a large fictitious acceleration.

To show its impact on the cost function, we utilize the following formula that turns the area integral of the $L^2$ norm of the convective acceleration in potential flows to a line integral over the boundary:

$$\frac{1}{2} \int \int |u \cdot \nabla u|^2 dx dy = \frac{1}{32} \oint \left( \nabla |u|^4 \right) \cdot n ds, \tag{10}$$

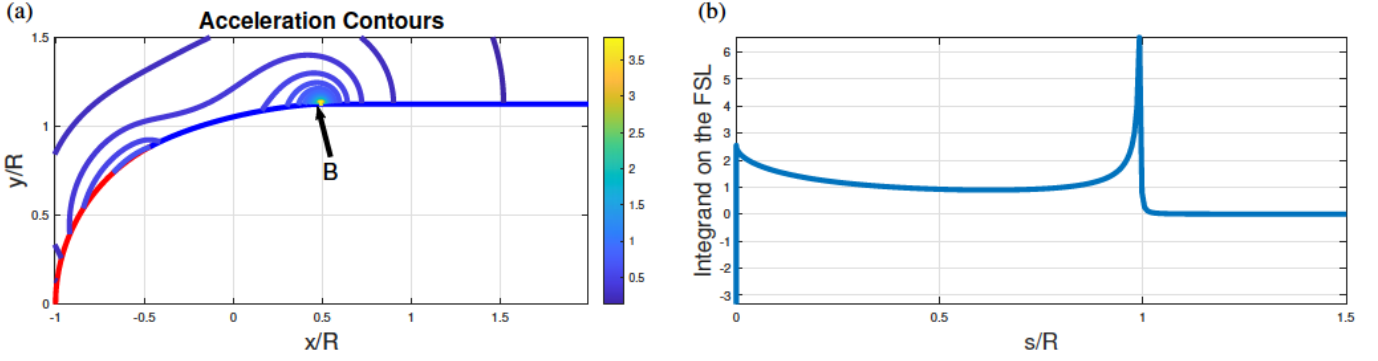

Figure 6: Inappropriateness of Roshko's model for integration with the Principle of Minimum Pressure Gradient due to its introduction of a fictitious jump in the acceleration magnitude at point $B$, affecting the pressure gradient cost. (a) A fictitious jump in the contours of acceleration magnitude in Roshko's flow (corresponding to $\beta_s = 62°$) where the region around point $B$ experiences the largest acceleration in the entire field. (b) Variation of the cost integrand $(\nabla |u|^4) \cdot n$ along the free streamline, showing a non-physical large peak at point $B$.

where $n$ is the unit normal to the boundary and $ds$ is the length element along the boundary. For the problem at hand, $\nabla |u|^4$ is non-zero only on two boundaries: the cylinder and the free streamline (FSL). Therefore, the steady Appellian $S$ is simply the sum of two line integrals. Figure 6 (b) shows the variation of the cost integrand $(\nabla |u|^4) \cdot n$ along the FSL which shows a non-physical large peak at the point B. It is interesting that the flow field at B is very smooth, giving no clue about a significant jump in acceleration at this point. This behavior emphasizes the importance of acceleration considerations when developing theoretical models.

In conclusion, while Roshko's model provides an elegant means of theoretical modeling of separating flows using ideal-flow theory, it is not appropriate for integration with the PMPG because of two main issues:

1. It lacks robustness due to the high sensitivity of flow features and the Appellian cost to small changes in the free parameters of the model, which presents a formidable challenge in formulating the problem in an optimization framework.

2. It does not accurately capture the physics of the acceleration field, which is the primary concern of the PMPG. Any numerical optimizer attempting to obtain the least-curvature solution (i.e., minimizing the $L^2$ norm of acceleration) will be misled by the fictitious acceleration induced by the model at B.

Therefore, we choose to integrate the PMPG with a simpler and more robust theoretical model of the separating flow over a cylinder, recently developed by [MM24]. This study will be the focus of the next section. That said, Roshko's model can still yield interesting insights into separation from the perspective of the PMPG, as discussed below.

If the least-curvature solution within Roshko's family cannot be determined, we can at least compare their Appellian costs to that of the attached flow. Consider the well-known attached, potential flow over a cylinder [Kar66, CI02, ST79]:

$$W(z) = U_\infty \left(z + \frac{R^2}{z}\right),$$

where $R$ is the raidus of the cylinder. One can simply derive an analytical expression for its Appellian cost as

$$S_\text{attached} = \frac{3}{2} \pi \rho U_\infty^4.$$

Interestingly, it is independent of $R$. On the other hand, recall any reasonable solution from Roshko's family. For example, consider Solution 4 in Table 2, which corresponds to $(k, \beta_s) = (1.4, 80°)$. Its associated cost is calculated as

$$S_\text{separating} = 0.93 \rho U_\infty^4,$$

which is five times smaller than the cost of the attached flow. In fact, it is not only the cost of Solution 4 that is smaller than that of the attached flow; almost all members in Roshko's family of candidate flows are more economic than the attached flow in terms of the curvature cost $S$.

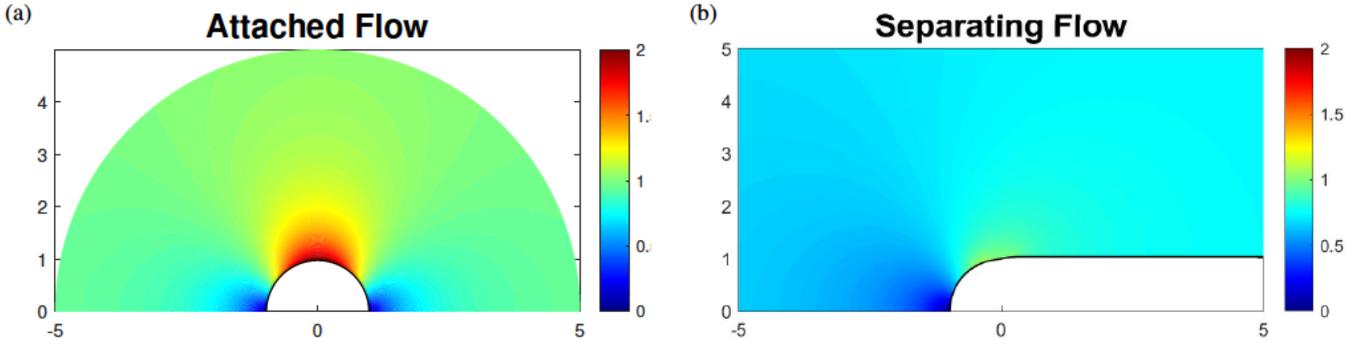

Figure 7: Contours of velocity magnitude in the cases of: (a) Attached flow, (b) Separating flow (Solution 4 in Roshko's family presented in Table 2).

The two flow fields (the attached flow and Solution 4) are juxtaposed in figure 7. It is known that, at high Reynolds numbers, Nature picks the separating configuration. This separation is typically attributed to boundary layer dynamics subject to an adverse pressure gradient. Here, the PMPG provides an alternative perspective that is not necessarily contradicting the conventional wisdom. The main philosophy of the PMPG is to turn the fluid mechanics problem into minimization. From this perspective, the adverse pressure gradient and its concomitant separation occur because it is more economic in terms of the cost function: the Appellian $S$. The demanded curvature is much higher if the flow closes behind the cylinder. It is simply easier for the flow to separate. In other words, larger pressure gradient forces would be needed to satisfy the continuity constraint for the attached case than in the separating case, which contradicts Gauss' principle of least constraint: the physically correct motion is the one that requires the smallest constraint force.

## 5 Matheswaran and Miller's Model

[MM24] have recently developed a very simple model for the separating flow behind a cylinder, by adding a source of unknown strength $Q$ to the stream function of the classical potential flow around a cylinder. The complex potential and complex velocity functions are then written as:

$$W(z) = U_\infty \left(z + \frac{a^2}{z}\right) + \frac{Q}{2\pi} \ln(z - z_s), \tag{11}$$

$$\frac{dW}{dz} = u - iv = U_\infty \left(1 - \frac{a^2}{z^2}\right) + \frac{Q}{2\pi(z - z_s)}, \tag{12}$$

where $a$ is the radius of the *initial* cylinder, and $z_s$ is the location of the point source. Just by adding this point source, the resulting flow looks like a separating flow over a slightly larger cylinder (of radius $R > a$), as shown in figure 8 (a).

The model then has three free parameters: $a$, $Q$, and $z_s$. To preserve the location of the forward stagnation point at $z = -R$ (taking unit radius $R = 1$), Matheswaran and Miller enforced the following condition:

$$\hat{Q} = 4\pi(1 - a^2), \tag{13}$$

where $\hat{Q} = \frac{Q}{U_\infty}$. Moreover, symmetry dictates that the point source must lie on the real axis (i.e., $z_s$ is a real number). Matheswaran and Miller opted to take $z_s = 1$. Finally, they enforced the separation condition: At a given separation angle $\beta_s$, the velocity must have a magnitude of $kU_\infty$ for an arbitrarily given $k \geq 1$. This condition is written mathematically as

$$\left|\frac{dW}{dz}\left(z = e^{i(\pi - \beta_s)}\right)\right| = kU_\infty,$$

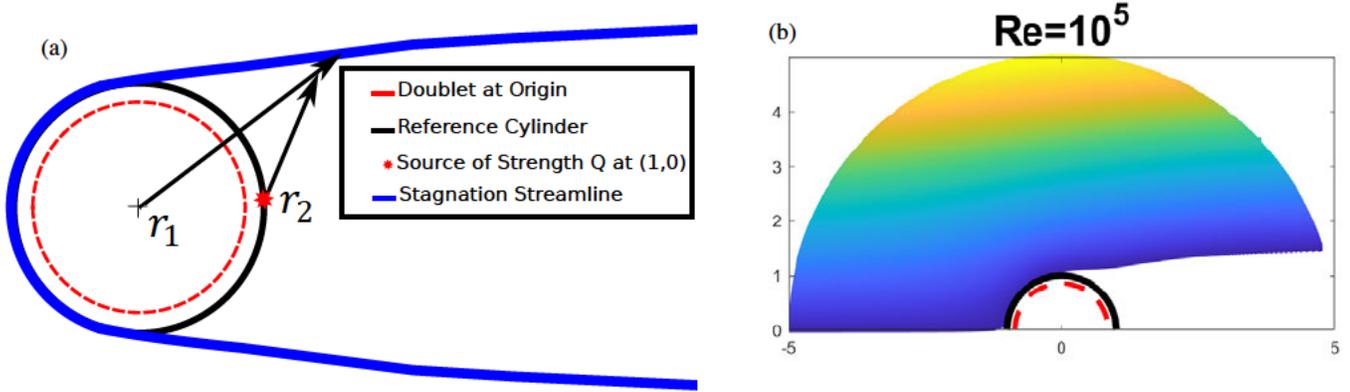

Figure 8: (a) Schematic of the model of [MM24] for the separating flow behind a cylinder, constructed by adding a source of unknown strength $Q$ (the red star) to the stream function of the classical attached potential flow. (b) A sample flow from [MM24]'s model at Re=$10^5$, showing a reasonable near-wake configuration for the separating flow over a cylinder.

which results in the following equation for the normalized strength $\hat{Q}$ of the point source:
$Q$ is then determined by solving the following quadratic equation:

$$\frac{\hat{Q}^2}{4\pi^2}\left[\left(\frac{z_1}{2}+z_3\right)^2+\left(z_4+\frac{z_2}{2}\right)^2\right]+\frac{\hat{Q}}{\pi}\left[(1-z_1)\left(\frac{z_1}{2}+z_3\right)-z_2\left(z_4+\frac{z_2}{2}\right)\right] \quad + \\
+\left[z_2^2+(1-z_1)^2-k^2\right] \qquad\qquad = 0, \tag{14}$$

where the constants $z_1$-$z_4$ are given by:

$$\begin{aligned}
z_1 &= \mathrm{Re}\left(\frac{1}{e^{2i\beta_s}}\right) = \cos(2\beta_s), \\
z_2 &= \mathrm{Im}\left(\frac{1}{e^{2i\beta_s}}\right) = \sin(2\beta_s), \\
z_3 &= \mathrm{Re}\left(\frac{1}{e^{i\beta_s}-1}\right) = -\frac{1}{2}, \\
z_4 &= \mathrm{Im}\left(\frac{1}{e^{i\beta_s}-1}\right) = -\frac{1}{2}\tan\left(\frac{\beta_s}{2}\right).
\end{aligned} \tag{15}$$

For a given combination of $(k, \beta_s)$, equation (14) results in two roots for $\hat{Q}$, which can then be used to determine $a$ form equation (13). The root that results in a real value of $a$ is picked. That is, Matheswaran and Miller's model has two free parameters, $k$ and $\beta_s$, to fully describe the flow, which aligns with our analysis of Roshko's model, where $k$ alone was found insufficient to predict the separation point $\beta_s$. To close their model, Matheswaran and Miller relied on experimental data and empirical relations to determine both parameters at a given Reynolds number.

Collecting experimental data from different sources [GWQH12, MY16, NT01, Nor87, NS87, WOS+11], they constructed the following empirical relation for $k$ in terms of the Reynolds number $Re$:

$$k = 1.51 - 0.211 e^{(-0.000121 Re)}. \tag{16}$$

In addition, they relied on Jiang's empirical formula for the time averaged separation angle $\bar{\beta}_s$ in the sub-critical regime $300 \leq Re \leq 3 \times 10^5$ as a function of $Re$ [Jia20]:

$$\bar{\beta}_s = 78.8 + \frac{505}{\sqrt{Re}} \quad \left(270 \leq \mathrm{Re} \leq 10^5\right). \tag{17}$$

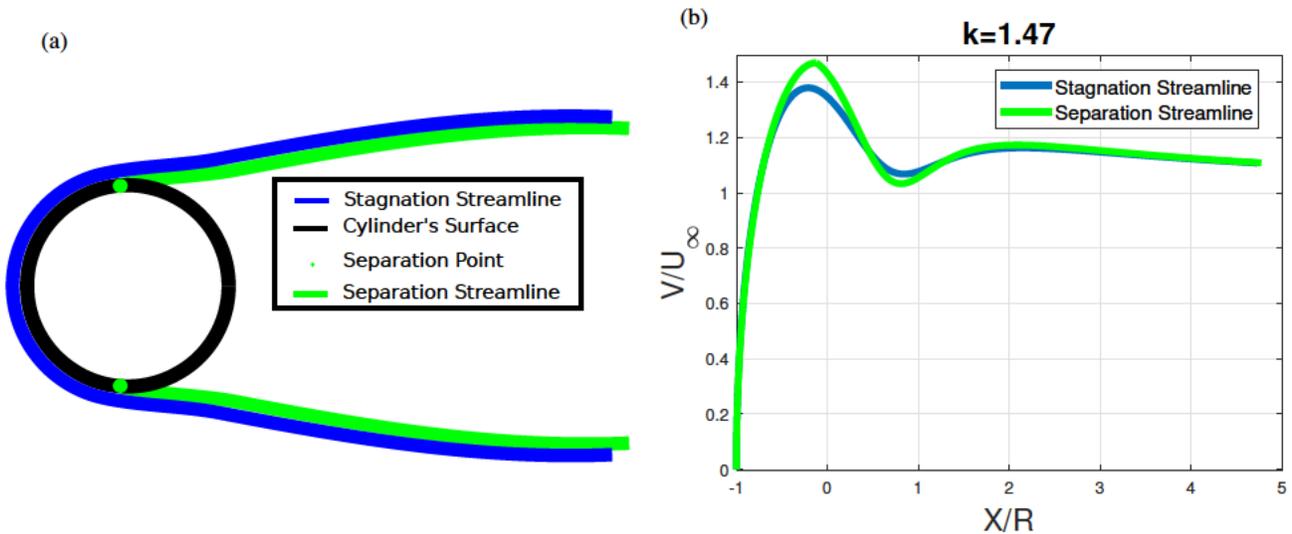

Figure 9: [MM24]'s model characteristics. (a) The stagnation streamline coincides with the cylinder's surface only at the forward stagnation point and instantly diverges, implying that the separation point on the cylinder's surface, $z = e^{i\beta_s}$, where the condition (14) is enforced, does not belong to the stagnation streamline. (b) Speed variation along both the separation and stagnation streamlines causes inaccuracy in the constant speed assumption of the free streamline theory.

Table 3: Comparison of the separating flow models of [Ros54] and [MM24].

| Criteria | [Ros54] | [MM24] |
|---|---|---|
| Mathematical Rigor | Stagnation & separation streamlines are the same. | Stagnation & separation streamlines are different. |
| Simplicity | ✗ | ✓ |
| Robustness | Highly Sensitive to slight variations in the model parameters. | Robust to small changes. |
| Accuracy | The velocity is constant along the free streamline. | The velocity varies along the free streamline. |
| Fictious Jump Acceleration | ✗ | ✓ |

Figure 8 (b) shows a sample flow from Matheswaran and Miller's model at $Re = 10^5$. The model appears to produce a reasonable near-wake configuration for the separating flow over a cylinder with admirable simplicity, in comparison to Roshko's model, which required four conformal transformations across five planes to produce a similar flow field. However, the model of Matheswaran and Miller suffers from a serious drawback:

The stagnation streamline coincides with the cylinder's surface only at the forward stagnation point and instantly diverges from the surface, unlike the free streamline in Roshko's model which coincides with the cylinder's surface all the way up to the separation point. This implies that the separation point on the cylinder's surface, $z = e^{i\beta_s}$—where the condition (14), which ensures that the speed is $kU_\infty$, is enforced— does not belong to the stagnation streamline. Figure 9 (a) shows the stagnation streamline (which passes through the forward stagnation point), and the separation streamline (i.e., the streamline passing through $z = e^{i(\pi-\beta_s)}$). In Roshko's model, they coincide, resulting in a mathematically rigorous flow field. In Matheswaran and Miller's model, they are different, causing an inaccuracy in the speed along the "free streamline", as shown in Figure 9 (b), and also leading to some physical violations in the wake region; i.e., below the stagnation streamline). However, it should be emphasized that, in these simplistic models, the wake region is typically ignored. That is, the model validity should be assessed only above the stagnation streamline, while the wake region may be assumed stagnant, as traditionally done in various models of the free streamline theory [Kir69, Ros54, Ria21, Bro23].

Table 3 summarizes the pros and cons of Roshko's model and Matheswaran & Miller's model. In conclusion, while the latter model suffers from some drawbacks, its robustness and simplicity make it quite appealing for integration with the PMPG in an optimization framework, aiming to predict the separation angle without the need to explicitly model the boundary layer dynamics. This study will be detailed in the next section.

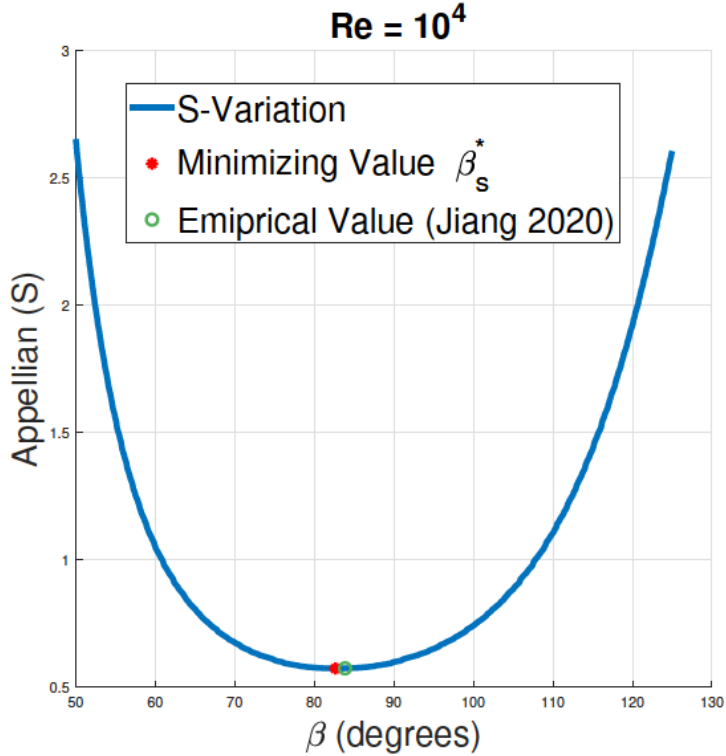

Figure 10: Variation of the pressure gradient cost (i.e., the Appellian in this inviscid case) with the separation angle $\beta_s$ at $Re = 10^4$ showing a unique value $\beta_s^* = 83.7°$ that minimizes the Appellian $S$ that is very close to the empirical prediction given by equation (17) from [Jia20] ($\beta_{\text{Empirical}} = 83.85°$).

## 6  Prediction of the separation angle $\beta_S$ using the PMPG

The strength of the PMPG is that it turns the fluid mechanics problem into a pure minimization problem, which makes it very convenient for theoretical modeling. For example, fluid mechanicians can utilize their understanding of the flow physics to develop a theoretical model that captures the important physics of the problem, typically with some free parameters. That is, the velocity field $u$ is written as $u = u(x; P_1, \ldots, P_n)$, where $P_1, \ldots, P_n$ are the free parameters of the model. Indeed, the velocity field in both Roshko's model [Ros54] and Matheswaran & Miller's model [MM24] can be expressed in the exact same form: $u = u(x; k, \beta_s)$. The PMPG can then be used to close the model by determining the values (or dynamics) of the free parameters via minimizing the cost (9). In the current formulation, we will focus on determining the separation angle $\beta_s$ only, since the wake parameter $k$ is related to the wake dynamics, which are not modeled in either Roshko's model or Matheswaran & Miller's model. As such, the fluid mechanics problem is formulated as the following one-dimensional optimization problem:

$$\min_{\beta_s} \quad S(\beta_s; k) = \frac{1}{2}\rho \int_{\mathcal{D}} |u_{MM}(x;\beta_s;k) \cdot \nabla u_{MM}(x;\beta_s;k)|^2 \, dx,$$

where $u_{MM}(x;\beta_s;k)$ is the Matheswaran-Miller (MM) flow field with $\beta_s$ being a free (unkown) parameter and $k$ is given, say from the empirical relation (16). So, for a given $k$, the Appellian cost $S$ is function of one variable: the separation angle $\beta_s$. It is then tempting to determine the value of $\beta_s$ that minimizes $S$, if there is any.

Figure 10 shows the variation of the pressure gradient cost (i.e., the Appellian in this inviscid case) with the separation angle $\beta_s$ at $Re = 10^4$. Interestingly, there is a unique value $\beta_s^* = 83.7°$ that minimizes the Appellian $S$. Moreover, the minimizing value is very close to the empirical prediction given by equation (17) from [Jia20] ($\beta_{\text{Empirical}} = 83.85°$). This accurate prediction of $\beta$ by minimizing the Appellian is not only achieved at $Re = 10^4$. The deviation between $\beta_s^*$ and $\beta_{\text{Empirical}}$ is less than 5% over the Reynolds number range between $10^4 - 10^5$, which is the same range where the flow characteristics are independent of $Re$ [Zdr97]. At smaller Reynolds number, , there is a considerable dependence of the flow characteristics on Reynolds number due to viscous effects, which are completely ignored by the adopted model of [MM24]; although the cost (8) can account for viscous effects, the ideal-flow representations of [Ros54] and [MM24] yield zero viscous stresses.

It is remarkable that, with such a simplistic model, the PMPG formulation enables a reasonable prediction of the separation angle without explicitly modeling boundary layer dynamics—a task once deemed impossible. This quantitative result aligns with the philosophy of the PMPG, mentioned earlier: while conventional wisdom attributes flow separation to adverse pressure gradient in the boundary layer, the PMPG asserts that both the adverse pressure gradient and separation develop to minimize the pressure gradient cost over the whole domain. Any other separating or non-separating solution would lead to an unnecessarily larger pressure gradient to maintain continuity. Equivalently, in this case, it would produce an unnecessarily larger curvature, as shown in figure 10.

Note that the Matheswaran & Miller model, which underpins the current analysis, was developed for the subcritical regime ($Re < 10^5$), where the boundary layer is still laminar. In this regime, while the wake is turbulent, the transition point lies downstream of the separation point along the free shear layer [Wu72]. With this in mind, the above result has significant implications: laminar separation at high Reynolds numbers can be predicted using an inviscid argument and an inviscid criterion (i.e., minimum curvature), without explicitly modeling boundary layer dynamics.

This result perfectly matches the visionary statement by Prandtl in his pioneering paper [Pra04]: *"The most important practical result of these investigations is that, in certain cases, the flow separates from the surface at a point entirely determined by external conditions"* appears less surprising when viewed from a different perspective. Since the pressure gradient normal to the wall is typically neglected in a laminar boundary layer, as noted by Prandtl [SK61], the adverse pressure gradient within the boundary layer is dictated by the outer inviscid dynamics, which indeed evolves to minimize the total curvature (i.e., the Appellian) of the outer flow. It is noteworthy to mention that the viscous cost of the boundary layer is given by

$$\mathcal{A} = \frac{1}{2}\rho \int_0^L \int_0^{\delta(x)} \left[P_x^2 + P_y^2\right] dy dx,$$

where $P_x \equiv \frac{\partial P}{\partial x}$, and $(x, y)$ are the tangent and normal coordinates relative to the wall, respectively. Here, $\delta$ is the boundary layer thickness, and $L$ is its length. Recalling Prandtl's assumption for laminar boundary layers: $P_y = 0$, we have

$$\mathcal{A} = \frac{1}{2}\rho \int_0^L \left[P_x^2(x)\delta(x)\right] dx, \tag{18}$$

which implies that the *viscous* cost of a laminar boundary layer is determined by the tangential pressure gradient $P_x$ at the edge of the boundary layer, which is governed by the outer inviscid dynamics; viscous effects enter through the boundary layer thickness $\delta$.

It is noteworthy to mention that figure 10 shows that the values of the appellian near the optimal separation angle $\beta_s^*$ appear relatively flat, which implies that if the flow separates at an angle slightly different from $\beta_s^*$, the penalty on the cost $S$ will not be significant. This suggests that minor variations in the flow field, caused by wake unsteadiness or viscous effects, may lead to considerable variations in the mnimizing angle, as the steady-state appellian does not appear to be heavily penalized in the neighborhood of $\beta_s^*$. This observation aligns with the well-established understanding that the separation angle does not remain fixed for each $Re$, but oscillates due to the unsteadiness of the flow field [WWY+04, MM67].

Finally, we must emphasize that full closure of the Matheswaran-Miller model was not possible; the value of $k$ remains to be determined form an external source. However, we believe that, with a more comprehensive model that explicitly accounts for the wake dynamics, one may be able to determine $k$, at least in the range $10^4 - 10^5$, by applying the unsteady version of the PMPG. This endeavor is left for future work.

## 7 Conclusion

In this paper, we performed two main studies for the separating flow over a circular cylinder using the free streamline theory. In this theory, the separating flow is modeled as an ideal flow with sheets of discontinuity that represent the separating shear layers. The theory provides a reasonable picture for the flow field in the near wake region, but with some free/undetermined parameters.

In the first study, we considered Roshko's model (Roshko 1954) of the free streamline theory. It is known that this model performs very well for cases with fixed separation (e.g., flat plate and wedge), but fails to predict the correct separation angle over curved surfaces. Roshko attributed the discrepancy in the latter case to the condition of matching curvature, which asserts that the curvature of the separating streamline at the separation point must match that of the cylinder. We studied the effect of matching curvature on Roshko's model and found that it is not the real culprit for the failure of Roshko's model to capture the separation angle. In fact, we discovered that Roshko's model is non-unique. There are solutions within Roshko's family of separating flows that satisfy the matching curvature condition while separating at angles that coincide with experimental measurements. We also justified the condition of matching curvature on physical grounds using the Principle of Minimum Pressure Gradient (PMPG), which asserts that an incompressible flow evolves from one instant to another in order to minimize the total magnitude of the pressure gradient over the domain. For the case of an ideal flow model, such as Roshko's, the pressure gradient cost reduces to the $L^2$-norm of the convective acceleration, which is a measure of curvature—a reminiscent of Hertz' principle of least curvature. We showed that if the condition of matching curvature is not satisfied, a significant jump in curvature and acceleration occurs at separation, which is non-physical according to the PMPG or least curvature.

In the second study, we focused on the subcritical regime of $Re = 10^4 - 10^5$ where the flow characteristics are known to be fairly independent of $Re$. This behavior suggests that we attempt to predict separation without modeling the boundary layer—a task typically considered impossible. To achieve this goal, we adopted a recently developed model for the free streamline theory by [MM24] that is simpler and more robust than Roshko's model. This model has the separation angle as a free parameter. Relying on the PMPG to select—from the Matheswaran-Miller family of separating solutions—the solution with least pressure gradient cost (equivalently least curvature). Interestingly, the obtained separation angles over the range $Re = 10^4 - 10^5$ closely match experimental measurements. It is known that separation occurs because of adverse pressure gradient n the boundary layer. This result provides an alternative perspective: Both the separation angle and its concomitant adverse pressure gradient develop to minimize curvature in the outer flow. Any other flow candidate will exhibit more curvature and require an unnecessarily larger pressure gradient force to maintain continuity. This result has significant implications: laminar separation at high Reynolds numbers may be predicted using an inviscid argument and an inviscid criterion (i.e., minimum curvature), without explicitly modeling boundary layer dynamics. Interestingly, this result matches Prandtl's visionary statement in his seminal paper [Pra04]: "*The most important practical result of these investigations is that, in certain cases, the flow separates from the surface at a point entirely determined by external conditions*"